\documentclass[doublecol]{epl2} 
\usepackage{graphics}
\usepackage{epsfig}
\usepackage{amssymb}
\usepackage{subfigure}
\usepackage{pifont}

\pdfoutput=1

\title{Active dry granular flows: rheology and rigidity transitions}
\shorttitle{Active dry granular flows: rheology and rigidity transitions}

\author{Anton Peshkov\inst{1,2} \and Philippe Claudin\inst{1} \and Eric Cl\'ement\inst{1} \and Bruno Andreotti\inst{1}}
\shortauthor{A. Peshkov \etal}
\institute{
\inst{1}Physique et M\'ecanique des Milieux H\'et\'erog\`enes, PMMH UMR 7636 ESPCI -- CNRS -- Univ.~Paris-Diderot -- Univ.~P.M.~Curie, 10 rue Vauquelin, 75005 Paris, France.\\
\inst{2}{Departamento de F\'isica, Facultad de Ciencias F\'isicas y Matem\'aticas, Universidad de Chile, Av. Blanco Encalada 2008, Santiago, Chile.}
}


\abstract{
The constitutive relations of a dense granular flow composed of self-propelling frictional hard particles are investigated by means of DEM numerical simulations. We show that the rheology, which relates the dynamical friction $\mu$ and the volume fraction $\phi$ to the inertial number $I$, depends on a dimensionless number $\mathcal{A}$, which compares the active force to the confining pressure. Two liquid/solid transitions -- in the Maxwell rigidity sense -- are observed. As soon as the activity is turned on, the packing becomes  an `active solid' with a mean number of particle contacts larger than the isostatic value. The quasi-static values of $\mu$ and $\phi$ decrease with $\mathcal{A}$.  At a finite value of the activity $\mathcal{A}_t$, corresponding to the isostatic condition, a second `active rigidity transition' is observed beyond which the quasi-static values of the friction vanishes and the rheology becomes Newtonian. For $\mathcal{A}>\mathcal{A}_t$, we provide evidence for a highly intermittent dynamics of this 'active fluid'.}

\begin{document}

\maketitle

%__________________________________
\section{Introduction}
\label{Introduction}

%%%%%%%%%%%%%%
\begin{figure}[t!]
\centerline{\includegraphics{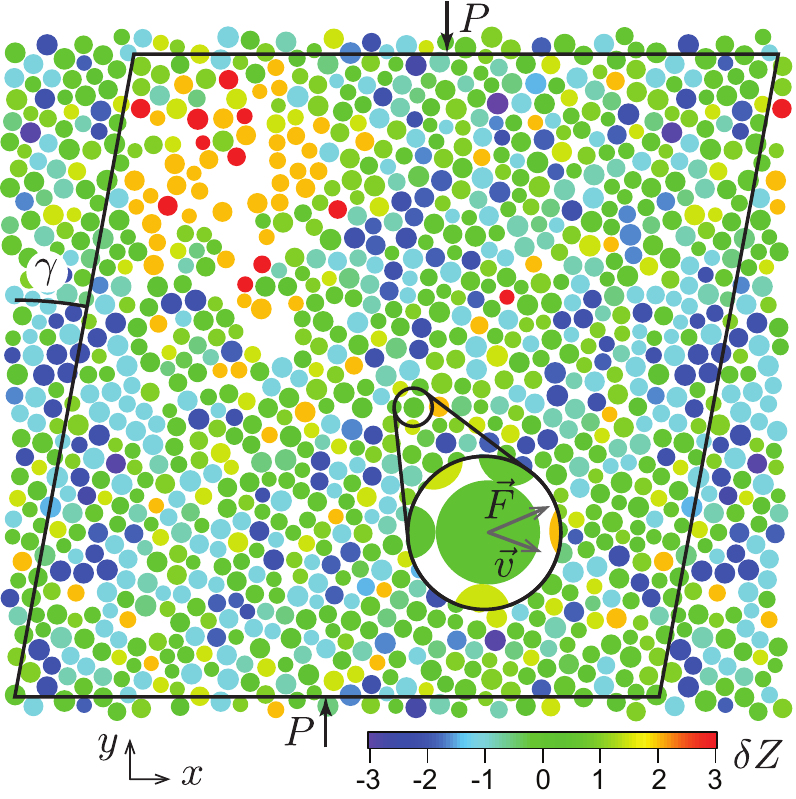}}
%\vspace{-5 mm}
\caption{Snapshot of the numerical system. Particles are submitted to contact forces from their neighbours, as well as an active force $\vec F$, oriented along the particle orientation unit vector and applied at the particle center. The system is submitted to a confining pressure $P$ and a shear rate $\dot \gamma$ in the $x$-direction. Here the grains are coloured according to the local value of the distance to isostaticity $\delta Z$ (Eq.~\ref{deltaZ}), see legend bar. This picture is from a simulation with an inertial number $I=10^{-4}$ and an active number $\mathcal{A}=0.6$. Spatio-temporal diagrams of Fig.~\ref{fig:SpatioTemporal} are computed from central ($y=0$) lines of such snapshots.}
%\vspace{-4 mm}
\label{fig:setup}
\end{figure}
%%%%%%%%%%%%%%

Self-locomotion displays fascinating collective properties originating from the interplay between individual self-propulsion and interactions among individuals in the group \cite{VZ12}. Since the pioneering work of Vicsek et al. \cite{VCBJCS95}, a large number of statistical physics models were created to render the emergence of collective motion, essentially based on an interplay between ordering processes reflecting microscopic interactions and disordering effects stemming from noise \cite{GC04}. However, these models based on kinetic equations for particle motion do not necessarily lead to a clear mechanical picture. In some instances such as collections of individuals in cognitive interactions, mechanical concepts may not be relevant to capture the essence of self-organisation. The mechanical perspective is however essential to understand, for example, the dynamics of externally agitated or self-propulsive granular systems \cite{NRM07,DDC10,BD16}, active colloids sedimentation \cite{PCBYB10}, the rheology of swimming micro-organisms suspensions \cite{GABJ08,SA09,RJP10,LGDAC15} or the collective dynamics of eukariotic cell forming tissues \cite{AHTMFW11,PRGMPLBS10}.

There has been recently considerable efforts to understand such systems in the light of statistical mechanics approaches and decipher the role of activity, i.e. the possibility to transfer energy and momentum from the microscopic level via self-propulsion of particles up to the macroscopic scale \cite{MJRLPRAS13}. Conceptually, these systems pose deep questions on the extension to active matter of equilibrium or close to equilibrium concepts such as effective temperature or pressure \cite{SBCKKT15,TF15, GTLYBBCB15}. It was found either from experiments or particle-based numerical simulations, that activity has a considerable influence on the macroscopic output as it may create new states of matter characterised by the emergence of dynamical clusters \cite{TCBPYB12, BBKLBS13,PSSPC13}, or large scale collective dynamics \cite{DCCGK04}. For active colloidal suspension, the onset of glassy dynamics  \cite{NCSD13,B14} or colloidal crystallisation process \cite{BSL12} is deeply affected by self-propulsion. The presence of autonomous swimmers such as motile micro-organisms suspended in a fluid also affects significantly the macroscopic rheology \cite{RJP10,LGDAC15} and the associated transport equation\cite{MJRLPRAS13,CFMOY08,GLM10}.

%__________________________________
\section{Dimensional analysis}
\label{DimensionalAnalysis}
In this letter, we investigate the extension of dry granular flow dynamics to the case where grains undergo self-propulsion. This is achieved in practice by exerting an external force of modulus $F$ to the particles, whose orientation rotates together with that of the particles. The rheology of the system is measured in a simple shear flow. Following conventional rheology, one may fix the volume fraction $\phi$ of particles and impose the shear rate $\dot \gamma$. Comparing inertial effects to the active force, one builds the dimensionless control parameter as
\begin{equation}
\Gamma = \frac{\rho^{1/2} d^{3/2} \dot\gamma}{ F^{1/2}} \, ,
\end{equation}
where $d$ and $\rho$ are the particle size and density. The rheology is then characterised by the shear stress $\tau$ or equivalently the viscosity $\eta=\tau/\dot \gamma$. In systems presenting a rigidity transition, it is more convenient to introduce the inverse of the viscosity, which is a fluidity. As the only parameters providing mass, length and time scales are $F$, $d$ and $\rho$, the dimensionless fluidity $G(\phi,\Gamma)$ is here defined as:
\begin{equation}
G = \frac{F^{1/2} \rho^{1/2} d^{1/2} \dot\gamma}{\tau} \, ,
\end{equation}
The very same rheology may, however, be expressed in a completely different way. Starting from the standard granular situation ($F \to 0$), the relevant energy scale is provided by  the confining pressure $P$, which is the control parameter conjugate of the volume fraction. The shear rate is then naturally rescaled to form the inertial number:
\begin{equation}
I  = \frac{ \rho^{1/2} d \dot\gamma}{ P^{1/2}} \, .
\end{equation}
Rescaling $F$ and the shear stress $\tau$ using $\rho$, $P$ and $d$, one defined the activity as:
\begin{equation}
\mathcal{A}=\frac{F}{Pd} \, ,
\label{ActiveNumber}
\end{equation}
and one forms the friction coefficient $\mu(I,\mathcal{A})$  \cite{Midi04,CEPRC05,JFP06,AFP13}, which replaces $G(\phi,\Gamma)$, as:
\begin{equation}
\mu=\frac{\tau}{P} \, .
\end{equation}
This formulation has proved to be very successful to capture the rheology of dense granular suspensions when $I$ is replaced by the viscous number $J$ \cite{BGP11}, and to understand the transition from the inertial to the overdamped viscous regime \cite{TAC12}. Recently this granular point of view was used to shed new light on the passage between the colloidal regime near the glassy transition and the viscous suspension regime near jamming \cite{TBKCCA15}. Here, we investigate its extension to active granular matter: we determine the constitutive relations and characterise the organisation processes associated with changes in the rheological response.

%__________________________________
\section{Numerical method}
\label{NumericalMethod}
The rheology of active dry grains under shear is investigated by means of numerical simulations (Discrete Element Method), using a modified LAMMPS code \cite{LAMMPS}. We consider a two-dimensional system (Fig.~\ref{fig:setup}), constituted of $N = 10^3$ circular particles labeled by an index $i$, with diameters $d_i$ randomly chosen in a flat distribution between $0.8\;d$ and $1.2\;d$, where $d$ is the mean particle size. Such a polydispersity is a compromise ensuring that the sample remains statistically homogeneous and does not crystallise when sheared. The mass density of the particles is denoted as $\rho$ \cite{note}. The system is submitted to a steady and uniform shear rate $\dot \gamma$ in the $x$-direction. We call $y$ the transverse direction. Periodic boundary conditions are used in both $x$ and $y$ directions (Lees-Edwards conditions), and the shearing is imposed by deformation of the numerical box. 

In the simulations, the contact between two particles $i$ and $j$ is modelled as harmonic springs with spring constants $k_n$ and $k_t$ (here we set $k_t = k_n/2$) for the normal and tangential components of the contact force, respectively $f_{ij}^n$ and $f_{ij}^t$. The contact is frictional with a coefficient $\mu_p=0.4$, and the Coulomb sliding condition ensures that $f_{ij}^t$ is never larger than $\mu_p f_{ij}^n$. Importantly, $k_n$ is large enough to be in the asymptotic rigid limit where results are independent of its value. In practice, simulations are performed with $k_n\simeq 10^4\,P$. The coefficient of restitution of the particles is $e=0.1$. We checked that the results are qualitatively independent of the values of $\mu_p$ and $e$.

The particle $i$ is not only submitted to contact forces ${\vec f}_{ij}$ from its neighbours $j$, but also to an `active' force $F {\vec e}_i$ aligned along its orientation unit vector ${\vec e}_i$, which is attached to the particle and which rotates due to the torque generated by the contact forces. The dynamics of the grains, characterised by their position, velocity $\vec u$ and angular velocity $\vec \omega$, is governed by
\begin{eqnarray}
m_i \frac{{\rm d} {\vec u}_i}{{\rm d} t} & = & \sum_j {\vec f}_{ij} + F {\vec e}_i,
\label{acc} \\
M_i \frac{{\rm d} {\vec \omega}_i}{{\rm d} t} & = & \frac{d}{2} \sum_j {\vec n}_{ij} \times {\vec f}_{ij},
\label{accang}
\end{eqnarray}
where $m$ and $M$ are respectively the mass and the moment of inertia of the particle. ${\vec n}_{ij}$ is the contact direction between particles $i$ and $j$. These equations are integrated following a velocity-Verlet integrator.

%%%%%%%%%%%%%%
\begin{figure}[t!]
\centerline{\includegraphics{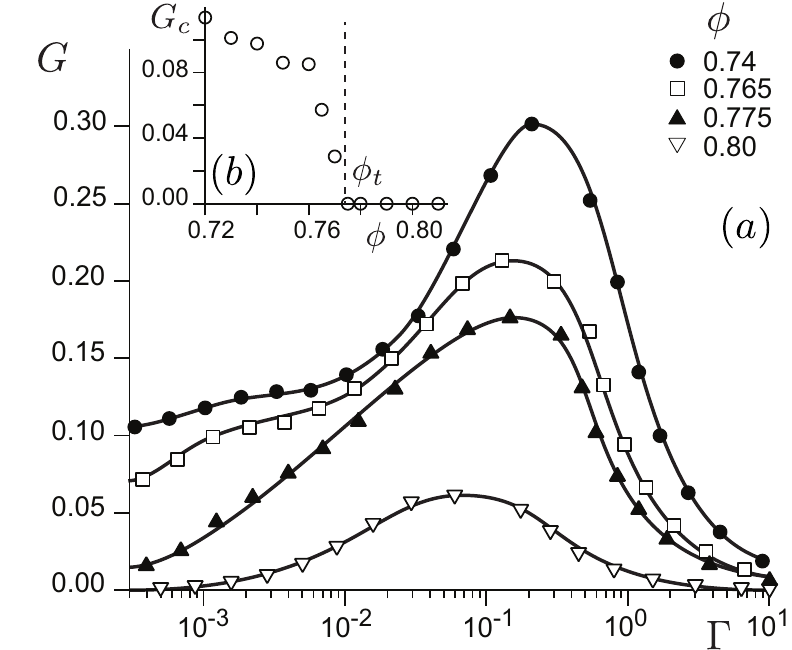}}
%\vspace{-5 mm}
\caption{Rheology at constant volume fraction $\phi$. (a) Fluidity $G$ as a function of the dimensionless shear rate $\Gamma$ for different values of $\phi$ (legends). Data points computed by interpolation of the pressure-controlled data set. (b) Extrapolated value of $G$ in the quasi-static limit $\Gamma \to 0$ as a function of $\phi$. Vertical dashed line: visualisation of the critical value $\phi_t \simeq 0.774$.}
%\vspace{-4 mm}
\label{fig:GK}
\end{figure}
%%%%%%%%%%%%%%

The contact stress components are measured as $\sigma_{\alpha\beta}=\frac{1}{2 S} \sum_{ij} {f}_{ij}^{\alpha} {r}_{ij}^{\beta}$ over the surface area $S$ of the cell, where ${\vec r}_{ij}$ is the branch vector between the centers of grains $i$ and $j$. The shear stress is $\tau=\sigma_{xy}$ and the confining pressure is given by $P=\sigma_{yy}$. Following the standard approach in dense granular flows \cite{Midi04,CEPRC05,JFP06,BGP11,TAC12}, we work at imposed homogeneous and constant confining pressure $P$. We have checked that in the limit of large systems, this is entirely equivalent to work at imposed volume fraction $\phi$. In a shear cell with an upper and a lower wall, the barostat directly acts on the dynamics of these walls along the $y$-axis. Here, because of the periodic boundary conditions, the pressure is controlled by the dilatation or shrinking of the simulation box along the $y$-direction. Using the standard Nos\'e-Hoover algorithm, the barostat dynamics follows a first order relaxation process, controlled by a relaxation time set to a small fraction of the shear time: $T_b = 3.2 \cdot 10^{-2}/\dot\gamma$. The numerical factor has been adjusted in the passive case ($\mathcal{A}=0$), in order to recover the well established behaviours of dry granular flows at both small and large shear rate. As there is no reference fixed wall, the centre of mass of the system is also free to move and we denote as ${\vec v}_{\rm cm}$ its velocity in a fixed referential. To be physically relevant, particle velocities are measured with respect to ${\vec v}_{\rm cm}$. In order to prevent the numerical divergence of $v_{\rm cm}$, a first order relaxation dynamics, similar to that of the barostat, is applied to the center of mass of the system to control its location (i.e. a `locostat'). It is controlled by a relaxation time here set to $T_l = 8 \cdot 10^{-4} \sqrt{P/\rho}/d$. Again, the numerical factor has been empirically adjusted and we have checked that it does not affect the results within a reasonable range.

Simulations are run for given values of the inertial number $I$ and of the active number $\mathcal{A}$, in the range $10^{-4}$ to $3.2 \cdot 10^{-1}$ for $I$ and $0$ to $1$ for $\mathcal{A}$.  After the completion of the relaxation transient, all macroscopical and microscopical parameters are averaged in the steady and homogeneous state, over a time of no less than $10^2/\dot\gamma$. The rheology is presented in the following section. The results obtained by measuring microscopic quantities, related to particle orientation and contacts are studied in next section.

%__________________________________
\section{Rheological curves}
\label{RheologicalCurves}
Fig.~\ref{fig:GK} shows the rheology $G(\Gamma)$ of the active granular system for different values of $\phi$. The fluidity presents a maximum (a minimum viscosity) located at values of $\Gamma$ slightly smaller than $1$. This indicates a cross-over between a shear thinning behaviour in the regime dominated by the active force and a shear thickening regime when inertia dominates. In the limit of vanishing shear rate $\Gamma$, the fluidity tends to a constant $G_c(\phi)$ which vanishes above a critical volume fraction $\phi_t$. Above $\phi_t$, the viscosity therefore diverges at $\Gamma \to 0$, which simply means that the system behaves as a yield stress fluid. Below $\phi_t$ the viscosity tends to a finite constant at vanishing shear rate. $\phi$ therefore triggers a rigidity transition which is specific to active granular material. Importantly, the shear thinning at small $\Gamma$ is simply the signature of the existence of a yield stress: if the viscosity diverges at vanishing $\dot \gamma$ then it must decay with $\dot \gamma$. Reciprocally, in the limit of large $\Gamma$, one recovers the shear thickening regime of rigid, passive grains for which dimensional analysis shows that the viscosity must scale as $d^2 \dot \gamma$. The existence of a minimal viscosity is therefore related to the choice of the parametrisation, which mixes the effect of shear rate and the effect of the active force.

%%%%%%%%%%%%%%
\begin{figure*}[t!]
\centerline{\includegraphics{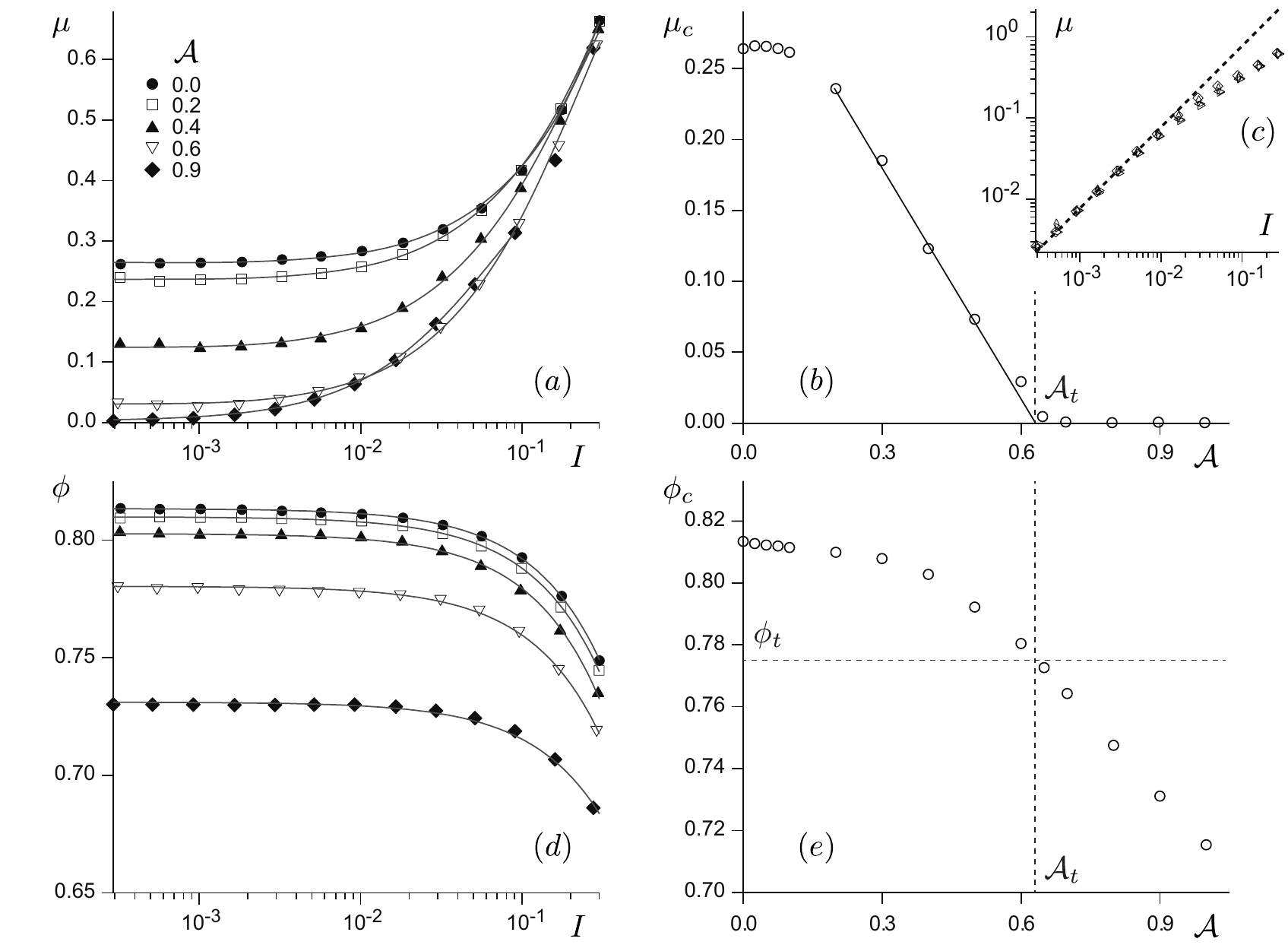}}
%\vspace{-5 mm}
\caption{Rheological constitutive relations at constant confining pressure $P$. (a) Dynamic friction $\mu$ as a function of $I$ for different values of $\mathcal{A}$ (legends). Symbols: numerical data. Solid lines: best fit by Eq.~\ref{muofI}. (b) Critical friction $\mu_c$ as a function of $\mathcal{A}$. Vertical dashed line: visualisation of the critical value $\mathcal{A}_t \simeq 0.635$. Horizontal dashed line: $\phi_t = \phi_c(\mathcal{A}_t )$. Solid line: linear fit vanishing at $\mathcal{A}_t$. Inset (c): $\mu(I)$ for $\mathcal{A} = 0.7$, $0.8$, $0.9$ and $1$ in log-log axes. Dashed line: linear scaling $\mu = 7.7 I$. (d) Volume fraction $\phi$ as a function of $I$ for different values of $\mathcal{A}$ (same legend as panel a). Solid lines: fit of Eq.~\ref{phiofI}. (e) Critical volume fraction $\phi_c$ as a function of $\mathcal{A}$. Error bars are smaller than the symbol size.}
%\vspace{-4 mm}
\label{fig:muphi}
\end{figure*}
%%%%%%%%%%%%%%

In order to disentangle these two effects, one needs to use the second possible parametrisation, where $I$ controls the shear rate independently of $\mathcal A$, which controls the active force. Although perfectly equivalent, the granular presentation of the rheology is better suited to the understanding. As illustrated in Fig.~\ref{fig:muphi}, the behaviour of the effective friction $\mu$ as well as that of the volume fraction $\phi$ continue to follow the empirical forms proposed for ordinary (passive) grains:
\begin{eqnarray}
\mu & = & \mu_c + \frac{\Delta\mu}{1+I_0/I},
\label{muofI}\\
\phi & = & \phi_c - b I,
\label{phiofI}
\end{eqnarray}
but where the parameters $\mu_c$, $\Delta\mu$, $I_0$, $\phi_c$ and $b$ now vary with $\mathcal{A}$. A central result is the behaviour of the effective friction at asymptotically low shear (Fig.~\ref{fig:muphi}b). $\mu_c(A)$ starts with a plateau $\mu_c^0 \simeq 0.26$ independent of the active number, up to $\mathcal{A} \simeq 0.2$. It then decreases down to zero at a value that can be linearly extrapolated to $\mathcal{A}_t \simeq 0.635$, above which it stays close to zero. As expected, the transition observed using the standard rheological approach survives the change of parametrisation. Instead of a critical volume fraction $\phi_t$, the activity appears to be the control parameter of the rigidity transition at vanishing $\dot \gamma$. At sufficiently small $I$, friction linearly increases with the inertial number as $\mu \sim \mu_c + a I$. The coefficient $a = \Delta \mu / I_0$ is on the order of $2$ in the passive case, and increases for larger $\mathcal{A}$. In the regime $\mathcal{A} > \mathcal{A}_t$, where $\mu \sim a I$ (Fig.~\ref{fig:muphi}c), the system is thus Newtonian with a  kinematic viscosity $\nu = \tau/(\rho\dot\gamma) \sim a d \sqrt{P/\rho}$ which becomes independent of the activity value. Note that this scaling is identical to that of a dilute gas, whose thermal velocity is $\sqrt{P/\rho}$ (the ideal gas relation), with a mean free path on the order of $ad$. This interpretation is consistent with a correlation length of the particle velocity found to be on the order of $10\,d$ (result not shown).

Activity also has an effect on the the critical volume fraction $\phi_c$, which decreases with $\mathcal{A}$ (Fig.~\ref{fig:muphi}d), i.e. making the system less dense for larger $\mathcal{A}$, as expected. The shape of the curve $\phi_c(\mathcal{A})$ is monotonous, starting at $\phi_c^0 \simeq 0.812$ in the passive case and decreasing with $\mathcal{A}$, with no particular change of behaviour is encountered around the value $\mathcal{A}_t$. Rather, one can observe a more negative slope above $\mathcal{A} \simeq 0.45$. The factor $b$ in (\ref{phiofI}) is found essentially constant, on the order of $0.2$. We have finally studied the normal stress difference $\delta \sigma = \sigma_{xx}/P-1$, focusing on the low shear regime (not displayed). This quantity remains close to zero in the whole range $0 \le \mathcal{A} \le \mathcal{A}_t$. Above the critical value of the active number, it increases in a linear fashion as $\delta \sigma = c(\mathcal{A} - \mathcal{A}_t)$, with $c \simeq 0.44$. As discussed below, this behaviour can be understood as a consequence of an orientation ordering of the grains above $\mathcal{A}_t$: they tend to align their orientation and velocity vectors along the $x$-direction.

%__________________________________
\section{Microscopic data}
\label{MicroscopicData}
\begin{figure}[t!]
\centerline{\includegraphics{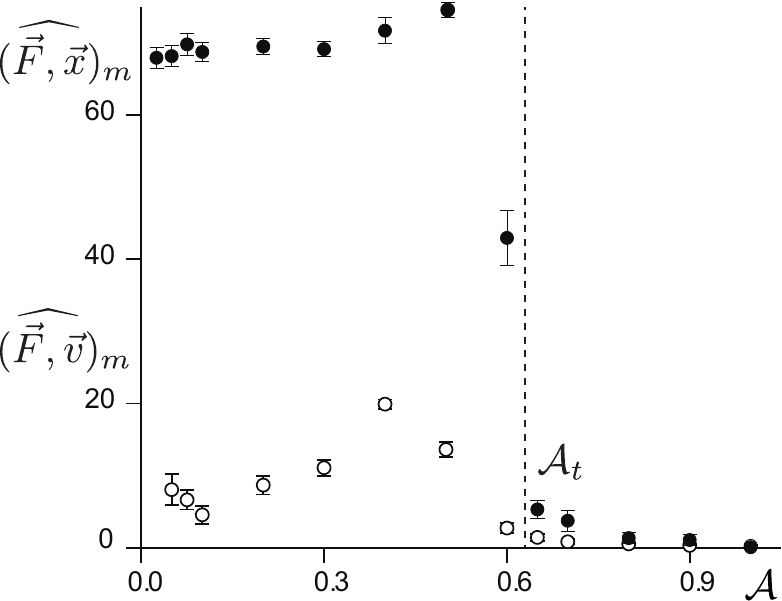}}
%\vspace{-5 mm}
\caption{Most probable angles (in degrees) of the particle orientation vector, which is aligned with the active force $\vec F$, with respect to the $x$ axis (filled symbols) and with respect to the orientation of the non-affine velocity (empty symbols), as functions of $A$. These data point correspond to the quasi-static limit $I \to 0$. At larger inertial numbers (not displayed), the angles $ (\widehat{\vec F,\vec x})_m$ and $ (\widehat{\vec F,\vec v})_m$ become almost independent of $\mathcal{A}$, with respective values around $60^\circ$ and $0^\circ$.}
%\vspace{-4 mm}
\label{fig:orientations}
\end{figure}

To complement the macroscopic rheological laws, we have studied several microscopic quantities: angular distribution of the particle orientation, which is aligned with the active force $\vec F$, with respect to the $x$-axis and with respect to the particle non-affine velocity $\vec v$ (Fig.~\ref{fig:setup}), as well as the coordination number $Z$. The non-affine velocity is computed as the difference between the particle velocity and the mean profile $\dot\gamma y \vec{x}$. Angle distributions typically present a sinusoidal shape, from which an amplitude and a most probable value, which depend on both inertial and activity numbers, can be extracted. In Fig.~\ref{fig:orientations}, we display the variations of the two most probable angles $(\widehat{\vec F,\vec x})_m$ and $ (\widehat{\vec F,\vec v})_m$ as functions of $\mathcal{A}$, in the quasi-static limit $I \to 0$. We see that these angles switch from finite and fairly constant values at small $\mathcal{A}$ to zero above the critical number $\mathcal{A}_t$. Activity then essentially does not affect the particle and velocity orientation, up to the transition where mean values of $\vec F$ and $\vec v$ both align along the $x$-direction. At larger $I$, the transition progressively disappears, and the two angles become independent of $\mathcal{A}$, where the vectors $\vec F$ and $\vec v$ align at some angle $\simeq 60^\circ$ with respect to  the $x$-direction. Importantly, these alignments are `weak' in the sense that the fluctuations around the mean values (i.e. the width of angle distributions) are large.

%%%%%%%%%%%%%%
%%%%%%%%%%%%%%

\begin{figure}[t!]
\centerline{\includegraphics{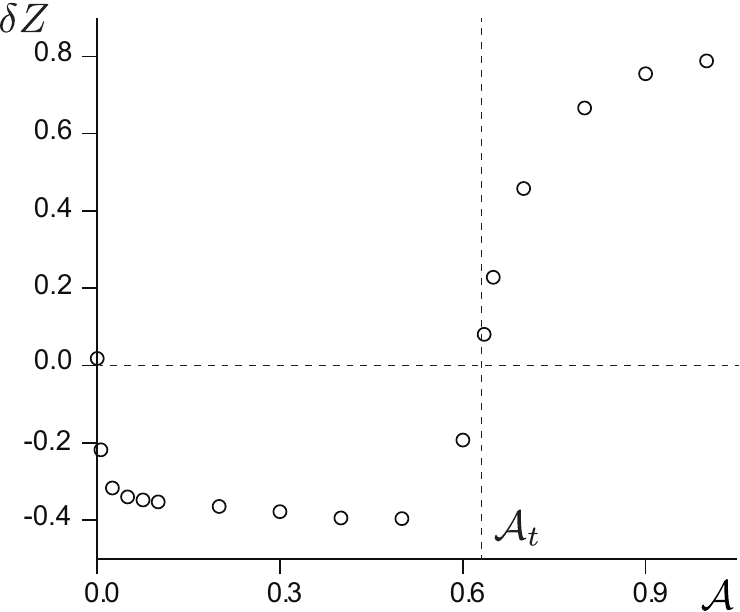}}
%\vspace{-5 mm}
\caption{Distance to isostaticity $\delta Z$ in the quasi-static limit $I \to 0$ as a function of $A$. Horizontal dashed line: $\delta Z=0$. Vertical dashed line: visualisation of the critical value $\mathcal{A}_t \simeq 0.635$. Error bars are smaller than the symbol size. At larger inertial numbers (not displayed), $\delta Z$ becomes almost independent of $\mathcal{A}$, with a value on the order of $1.5$ for $\mathcal{A}=1$.}
%\vspace{-4 mm}
\label{fig:deltaZ}
\end{figure}

We have also studied the connectivity of the grains to their neighbours. We call $Z$ the coordination number, i.e. the average number of contacts per grain. Following Maxwell rigidity criterion, we define the distance $\delta Z$ to isostaticity as the difference between the number of constrains and the number of degrees of freedom (the number of force components), here in the case of frictional particles in two dimensions as:
\begin{equation}
\delta Z = 3+\zeta-Z 
\label{deltaZ}
\end{equation}
where $\zeta$ is the fraction of sliding contacts \cite{AR07,vH10}.  A system is `solid', i.e. rigid, when hyperstatic ($\delta Z<0$) and `liquid' when hypostatic ($\delta Z>0$). It varies from $\delta Z = -3$ (grains with 6 non-sliding contacts) to $\delta Z = 3$ (grains with no contacts). In Fig.~\ref{fig:deltaZ} we show the variation of $\delta Z$ with respect to $\mathcal{A}$, in the quasi-static limit $I \to 0$. Starting from $\delta Z \simeq 0$ for the passive system, it quickly drops to a negative constant for the range $0<\mathcal{A}<\mathcal{A}_t$, and then switches to positive values above the critical activity. Crucially, this change of sign of $\delta Z$ also occurs at activity $\mathcal{A}_t$, where $\mu_c$ vanishes. At larger $I$ (not displayed), the transition progressively disappears and $\delta Z$ becomes almost independent of $\mathcal{A}$.

%%%%%%%%%%%%%%
%%%%%%%%%%%%%%

\begin{figure}[t!]
\centerline{\includegraphics{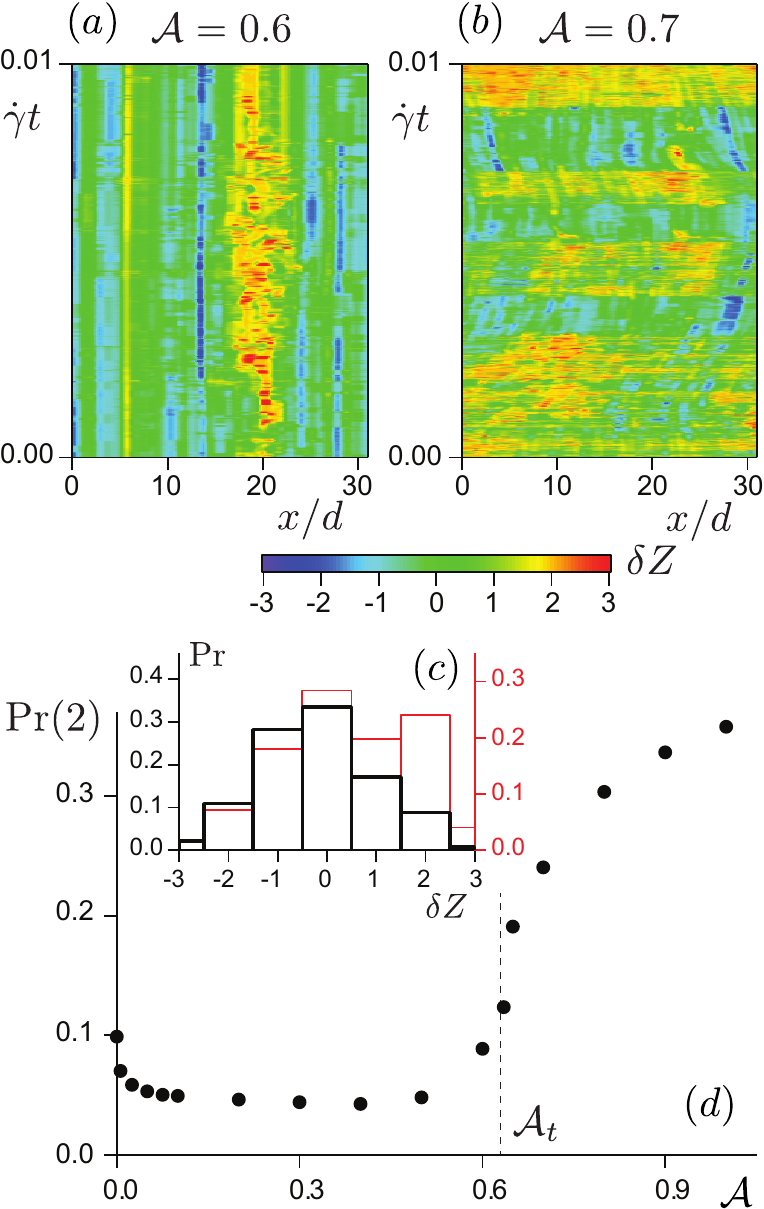}}
%\vspace{-5 mm}
\caption{Spatio-tempral diagrams constructed form horizontal central lines of snapshots such as that in Fig.~\ref{fig:setup}, for $\mathcal{A}=0.6$ (a) and $\mathcal{A}=0.7$ (b), i.e. below and above the transition $\mathcal{A}_t \simeq 0.635$. The same color code as in Fig.~\ref{fig:setup} is used. (c) Histograms showing the probability Pr of local $\delta Z$, i.e. computed on each grain. Thick black lines: $\mathcal{A}=0.6$ (left axis). Thin red lines: $\mathcal{A}=0.7$ (right axis). The scales on these two axes are different to collapse the left parts of the graphs, so that the excess of grains with $\delta Z \simeq 2$ above $\mathcal{A}_t$ is emphasised. (d) Evolution of the probability to find grains with $\delta Z \simeq 2$ when increasing $\mathcal{A}$. All of these data correspond to simulations at $I=10^{-4}$.}
%\vspace{-4 mm}
\label{fig:SpatioTemporal}
\end{figure}

The time-averaged curves of Figs.~\ref{fig:muphi}, \ref{fig:orientations} and \ref{fig:deltaZ} do not render the temporal fluctuations of the system. As the activity increases, the dynamics of the system becomes more intermittent. We can for example observe the formation of `bubbles' with no grains, as well as `clusters' of grains with large values of $\delta Z$ (often associated with these bubbles), as illustrated in the snapshot displayed in Fig.~\ref{fig:setup}. To further illustrate this dynamics, we show in Fig.~\ref{fig:SpatioTemporal}(a,b) spatio-temporal diagrams built from juxtaposition of horizontal lines extracted in such snapshots. We clearly see alternation of periods with positive and negative values of $\delta Z$ when the active number is above $\mathcal{A}_t$. This can be quantified in computing the histograms of local $\delta Z$. As evidenced in Fig.~\ref{fig:SpatioTemporal}(c), it changes from a single-peaked distribution dominated by negative values for $\mathcal{A}<\mathcal{A}_t$ to a wider double-peaked distribution for $\mathcal{A}>\mathcal{A}_t$, with a second maximum associated with positive values of $\delta Z$. This peak is around $\delta Z=2$, and its amplitude sharply increases above $\mathcal{A}_t$ (Fig.~\ref{fig:SpatioTemporal}(d)) in a way very similar to $\delta Z$ \emph{vs} $\mathcal{A}$ (Fig.~\ref{fig:deltaZ}). This corresponds to a larger number of grains with a single contact, which can be simply interpreted as self-propelled grains pushing their forward neighbours.

%%%%%%%%%%%%%%
%%%%%%%%%%%%%%

%%%%%%%%%%%%%%
\begin{figure}[t!]
\centerline{\includegraphics{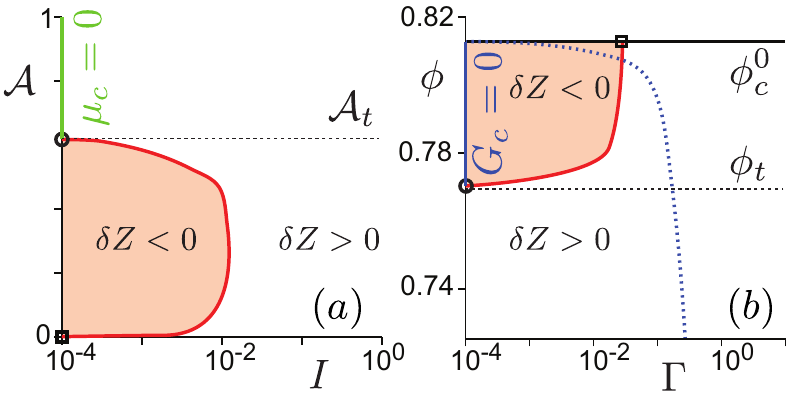}}
%\vspace{-5 mm}
\caption{(a) Phase diagram showing the hyperstatic ($\delta Z<0$) region (orange zone) in the constant pressure representation. Above $\mathcal A_t$, the friction coefficient in the limit $I \to 0$ vanishes (green line). (b) Corresponding diagram in the constant volume fraction representation. Above $\phi_t$, the fluidity in the limit $I \to 0$ vanishes (blue line). The dashed blue curve corresponds to a maximal fluidity (a minimal viscosity) for a given $\phi$.}
%\vspace{-4 mm}
\label{fig:IntertialEffects}
\end{figure}
%%%%%%%%%%%%%%

%__________________________________
\section{Discussion}
\label{Discussion}

We studied numerically how the rheology of hard and dry granular packing is modified by self-propulsive activity of the grains. Under a fixed confining pressure, activity decreases the resistance to shear and the average volume fraction. Importantly, the mechanism of friction reduction at play here does not rely on the mediation of a surrounding fluid as for active (dilute) suspensions but rather on a mechanism of collective organisation of the grains under shear. However, drawing on the unification of (dense) suspension and granular rheologies in the passive case \cite{BGP11,TAC12}, one can expect similar results for active dense granular suspensions in the viscous regime, and this direction of research is an obvious continuation of this work.

The dependence of the constitutive relations on the dimensionless parameter representing the active driving force, is non trivial and two rigidity phase transitions were observed. First, starting from isostatic conditions characterising a sheared granular packing near jamming, the packing gets into a state of 'active solid' in the sense that the number of contacts per grain becomes larger than the isostatic counting defining the rigidity transition. This first transition occurs as soon as the activity is turned on. It is reminiscent of the spontaneous clusterisation \cite{TCBPYB12,PSSPC13} or active jamming \cite{HFM11}, occurring for active packing in absence of shear (Fig.~\ref{fig:IntertialEffects}a). In the quasi-static limit, the rheology of this actively jammed state is characterised by a finite dynamical friction essentially controlled by the active number $\mathcal{A}$. In this state, the propulsive force is oriented in average \emph{off} the flow direction. The dynamical friction coefficient decreases continuously with activity and vanishes at a finite activity value ($\mathcal{A}_t \simeq 0.635$) (Fig.~\ref{fig:IntertialEffects}a). Equivalently, when the volume is controlled rather than the confining pressure, the fluidity at vanishing shear rate decreases with $\phi$ and vanishes above $\phi_t \simeq 0.774$ (Fig.~\ref{fig:IntertialEffects}b). The sensitivity of these critical values with system size and particle properties is to be systematically investigated.

This second transition occurs at the isostatic point and beyond it (for $\mathcal{A}>\mathcal{A}_t$ or $\phi<\phi_t$), the rheology is that of an effective Newtonian viscous liquid (Fig.~\ref{fig:IntertialEffects}). Interestingly, below or above the transition, the number of contacts in excess of the isostatic point varies continuously with the activity but the propulsion direction orientation seems to undergo a finite jump to almost zero in average. Even though the effective viscosity of this `active fluid' seems Newtonian, the contact number dynamics remains very intermittent alternating `solid' and `liquid' phases in the rigidity Maxwell sense. In the future it would be interesting to understand in depth this new type of transition and the nature of the active states of matter hence produced.

%___________________________________________________________________________
\acknowledgments
We thank M. Bouzid and M. Trulsson for discussions. AP is suported by Fondecyt 2015 Postdoctoral grant 3150406. BA is supported by Institut Universitaire de France. This work is funded by the ANR JamVibe and a CNES research grant.

%___________________________________________________________________________


\begin{thebibliography}{99}
\bibitem{note}
Note that in Refs.~\cite{Midi04,CEPRC05,JFP06} and in most of the papers dealing with the $\mu(I)$-rheology of granular flows, $\rho$ is rather defined as $m/d^2$, where $m$ and $d$ are the mean mass and diameter of the grains. This slight change of convention leads to a change by a factor $\sqrt{\pi}/2$ in the numerical values of $I$ for similar conditions.
\end{thebibliography}
\end{document}